\documentclass[12pt,preprint]{aastex}
\usepackage{amsmath,color,subfigure}

\begin{document}

\title{Population synthesis of young isolated neutron stars:
the effect of fallback disk accretion and magnetic field evolution}

\author{Lei Fu\altaffilmark{1,2} and Xiang-Dong Li\altaffilmark{1,2}}
\affil{$^{1}$Department of Astronomy, Nanjing University,
Nanjing 210093, China; lixd@nju.edu.cn}

\affil{$^{2}$Key laboratory of Modern Astronomy and Astrophysics
(Nanjing University), Ministry of Education, Nanjing 210093, China}

\begin{abstract}
The spin evolution of isolated neutron stars (NSs) is dominatd by their
magnetic fields. The measured braking indices of young NSs
show that the spin-down mechanism due to magnetic dipole radiation
with constant magnetic fields is inadequate.
Assuming that the NS magnetic field is buried by supernova fallback
matter and re-emerges after accretion stops,
we carry out Monte-Carlo simulation of the evolution of young NSs,
and show that most of the pulsars have the braking indices
ranging from $-1$ to 3. The results are compatible with the observational
data of NSs associated with supernova remnants. They also suggest that
the initial spin periods of NSs might occupy a relatively wide range.

\end{abstract}

\keywords{accretion, accretion disks $-$ pulsars: general $-$ stars: neutron
$-$ stars: magnetic fields}

\section{Introduction}
The spin evolution is one of the most outstanding problems
in pulsar astronomy. In the classical models the spin-down of
isolated pulsars is due to the energy loss caused by
magnetic dipole radiation, which is described as
\begin{equation}
I\dot{\Omega}=-\frac{B^2R^6\Omega^3}{6c^3},
\end{equation}
where $\Omega$, $\dot{\Omega}$, $I$, $B$, and $R$ are the angular
velocity and its derivative, the moment of inertia, the surface magnetic field
strength, and the radius of the pulsar, respectively, $c$ is the speed of
light. In realistic case the pulsar's spin-down may deviate from
that due to pure dipole radiation, and a more general power-law form
is adopted,
\begin{equation}
\dot{\Omega}=-K\Omega^n,
\end{equation}
where $K$ is a coefficient proportional to the spin-down torque,
and the power-law index $n$ is the
so-called braking index. For a constant $K$,
$n\equiv\Omega \ddot{\Omega}/\dot{\Omega}^2$.
The initial spin period $P_0$
of the pulsar can be obtained by using Eq.~(2) once the
real spin-down age and the historically averaged
braking index is known,
\begin{equation}
P_0=\left[P^{n-1}-(n-1)\dot{P}T/P^{2-n}\right]^{1/(n-1)},
\end{equation}
where $P(\equiv\Omega/2\pi)$, $\dot{P}$, and $T$ are the current
period, the period derivative, and the age
of the pulsar, respectively.

So far, due to the uncertainties in timing measurements (e.g.
timing noise and glitches) which usually dominate the relatively small
values of the second derivative of $\Omega$, it is impossible to
measure the stable braking indices for the majority of pulsars
\citep{kms94}. Only for those pulsars with relatively stable
long-term spin-down, the secular braking indices have been
measured. By using 23 year timing data \citet{lpg93} first obtained the
braking index of the  Crab pulsar (B0531$+$21) to be
$n=2.51(1)$, for which the time evolution of $\dot{\Omega}$ has
been monitored for more than 40 years. The braking index
of the Vela pulsar B0833$-$45 was measured by \citet{lpg96} to
be $n=1.4\pm0.2$, which is significantly less than the
value expected in the magnetic dipole radiation model, indicating possible
changes of the magnetic moment and/or the effective moment of inertia.
Currently the braking indices have been
measured in 13 pulsars \citep[see][and references therein]{esp13}.
Except PSRs J0537$-$6910 and B1757$-$24 which have negative $n$, other pulsars all have $0<n<3$.
The smallest one is $n=0.9\pm0.2$
for PSR J1734$-$3333 \citep{elk11}, which has a spin period of $P=1.17\,\rm s$ and
a large period derivative $\dot{P}=2.3\times 10^{-12}$ ss$^{-1}$.
The deduced magnetic field strength
$B\simeq 5 \times 10^{13}\,\rm G$ makes it among the highest $B$
pulsars, and similar to those of magnetars \citep{ozv12}. The low
$n$ has been attributed to an increase of the dipole component
of its magnetic field \citep{elk11}.

%Based on the value of $n$, \citet{elk11} suggested that the
%magnetic field of PSR J1734$-$3333 is growing, and if the physics of
%this process remains the same it will rank into
%magnetars within less than 30 kyr.
%\textcolor[rgb]{0.00,0.00,0.98}
%{However an alternate evolution channal is proposed by \citet{cea12} using
%a fallback disk with a mass $\sim 10^{-7}\,M_{\odot}$ and a NS with a constant dipole
%magnetic field of $\sim 10^{12}\,\rm G$ they reproduced the observed $P$, $\dot{P}$, $\ddot{P}$
%and $L_X$ of this source in an age of $\sim 3\times 10^{4}\,\rm yrs$.
%In their model the disk torque is exerted on the NS while the pulsar is switch on
%since the inner radius of the disk is at $R_{\rm LC}$.
%We noted that since the required strength of $B$ field response for the $\dot{P}$ in field increasing
%model is differed from the disk torque model about 1 order of magnitude
%, one would expect more accurate observations to determine the $B$
%field(e.g. the absorption line due to electron cyclotron resonance) to give a discrimination.
%However, whether a model include field evolution can reproduces the observed properties like
%fallback disk model with no field evolution involved needs more detailed numerical work to test.}

The possible reasons for the deviation of $n$ from 3 have been
studied extensively. They include, e.g.,
the pulsar wind in which the high-speed particles take
the angular momentum away from the pulsar \citep{mt77},
the distortion of the magnetic field from a pure dipole field
\citep[e.g.][]{bt10} and a oblique rotator in which the magnetic
axis is misaligned with the axis of rotation \citep[e.g.][]{cs06}.
In particular, to explain that most of the
measured values are less than 3, there are models invoking
magnetic field increase \citep[e.g.][]{br88} or fallback disk
assisted spin-down \citep[e.g.][]{mlr01,mph01,aay01,cl06}
in young pulsars.

The growth of magnetic field in young neutron stars (NSs)
may be caused by the re-emergence of the magnetic field
\citep[e.g.][]{mp96,gpz99,pg07,ho11,bpl12,pvg12}, which is buried due to the
hypercritical accretion after the supernova events \citep{rom90}.
The re-emergence timescale depends on the total amount of accreted
matter, which is $\sim 10^3-10^4$  yr for total accreted masses
$\sim 10^{-4}-10^{-3}\,M_{\odot}$.
Relatively weak magnetic fields ($\sim 10^{10}-10^{11}$ G)
have been measured in a few
young NSs in supernova remnants (SNRs)
named central compact objects (CCOs)
\citep[][and references therein]{gha13},
which might have experienced the field burial and re-emergence process \citep{ho13}.

Spin evolution of NSs with a surrounding fallback disk
originated from the SN ejecta
has been investigated by many authors \citep[e.g.][]{chn00,alpar01,mph01,eee09,yps12}.
In these studies the NS magnetic fields were usually
assumed to be constant,
neglecting their possible evolution during and after the disk accretion.
In this paper we use a Monte-Carlo method to calculate the evolution of
young NSs. We assume that there is a fallback disk around
all newborn NSs. Depending on the accreted mass the NS magnetic fields
decay to some extent, and turn back to increase after accretion stops.
By taking into account the disk accretion and field evolution simultaneously,
we model the evolution of the spin periods and the braking indices
of the NSs. We also compare the calculated distributions of various
parameters with those derived from observations of young NSs.

The structure of this paper is as follows. We introduce the fallback disk
model in Section 2, and describe the evolution of the magnetic field and
the spin period of NSs, which is coupled with fallback disk accretion
in Section 3. We present the calculated results of our population synthesis
in Section 4. Discussion and conclusions are given in Section 5.

%have Different from \citet{yps12}, in our model the disk
%torque is only exerts on NS while the pulsar is switch off and the disk evolution
%is performed in viscous or dynamical timescale, further more we consider the
%magnetic field evolution during and after the accretion which is based on
%previous numerical results. However the idea of using fallback disk and field evolution
%to explain the observed braking indexes is not new, our aim here is to
%exploring this idea systematically by numerical procedure.

\section{The fallback disk mass transfer}
\subsection{Time-evolution of a fallback disk}
%The Fallback model is widely assummed since several observational properties of NSs
%can be explained when using this theroy like the IR excess in isolated NSs\citep{phn00},
%the X-ray luminosity in anomalous X-ray pulsars(AXPs)\citep{phn00} and the
%discrepancies between the real and characteristic age of pulsars\citep{mlr01}.

The SN fallback material is assumed to form from the metal-rich ejecta of
core collapse SNe \citep{mic88,che89}.
The formation of a fallback disk requires that at least part of the fallback material
possesses sufficient angular momentum.
Compared with the accretion disks
in binaries, the lifetime of which may be comparable with the evolutionary timescale of
the binary or the donor star,
the fallback disk has much shorter duration.
The first possible detection of the fallback disk was made by \citet{wck06}, who
reported the mid-infrared counterpart of the anomalous X-ray
pulsar (AXP) 4U 0142$+$61,
and interpreted this as a passive dust fallback disk outside the pulsar's magnetosphere
heated by X-ray irradiation.
Here we are not concerned with the formation processes of the fallback disk
which is rather complicated, but focus on its time-evolution.

The time dependence of the mass transfer rate of
the disk and the size of a fallback disk was studied before
\citep[e.g.][]{mph01,ea03,eee09,clg90}. Based on the self-similar solution
of the standard thin disk \citep{ss73,pri81} the evolution of the mass
transfer rate at the outer annulus of the disk $\dot{M}_{\rm tr}$
and the outer disk radius $R_{\rm d}$ can be written in the following form
\begin{equation}
\dot{M}_{\rm tr}(t) = \dot{M}_0 \left(1+t/t_0\right)^{-\alpha},
\end{equation}
\begin{equation}
R_{\rm d}(t) = R_0 \left(1+t/t_0\right)^{2\alpha-2},
\end{equation}
where the power exponent $\alpha$ is $5/4$ for the opacity dominated by
bound-free absorption, $R_0$ and $\dot{M}_0$ are the initial radius and mass
transfer rate of the disk, respectively. $t_0$ is the timescale of disk formation,
and is usually taken to be the dynamical timescale for the
disk \citep{mph01},
\begin{equation}
t_{\rm d}\simeq 6.6\times 10^{-5} T_{\rm c,6}^{-1} R_{0,8}^{1/2}\; {\rm yr}.
\end{equation}
where $R_{\rm 0,8}$ and $T_{\rm c,6}$ are the initial outer disk radius
in units of $10^8$ cm and the temperature at $R_{\rm d}$ in units of
$10^6{\rm \,K}$ (taken to be $1$ in this paper), respectively. However, \citet{eee09}
showed that $t_0$ is more likely to be close to the viscous timescale in the disk,
\begin{equation}
t_{\rm v}\simeq 3.19\times 10^{-4} (M_0/10^{-4}M_{\odot})^{-3/7} R_{\rm 0,8}^{25/14}\; {\rm yr},
\end{equation}
where $M_0=\int_0^\infty \dot{M}_{\rm tr}(t){\rm d}t=\dot{M}_0 t_0/(\alpha-1)$
is the initial disk mass, or
\begin{equation}
t_{\rm v}\simeq 2.58\times 10^{-4} (\dot{M}_0/10^{25}{\rm \,g\,s^{-1}})^{-3/10} R_{\rm 0,8}^{5/4}\; {\rm yr}.
\end{equation}
In our calculation we assign the maximum
value of $t_{\rm d}$ and $t_{\rm v}$ to $t_0$.

Since the temperature of the disk always decreases with time, at the later stage
the outer regions of the disk may become neutral and passive \citep{mph01,eee09}.
However, it was argued that the irradiation by the NSs
may be able to keep the disk ionized and prohibit the transition from
an active to a passive disk \citep{aay01}. Further calculation showed that
the critical temperature $T_{\rm p}$ corresponding to the lowest ionization
fraction that can generate viscosity in the disk is as low as
$T_{\rm p}\sim 100\,\rm K$ \citep{ace13}. Thus in this work we
do not consider the neutralization process in the fallback disk.

\subsection{Mass transfer and accretion}

It is noted that the fallback disk accretion process is likely to be non-conservative.
For a NS with a typical mass of $M=1.4M_\odot$ the accretion rate ($\dot{M}_{\rm acc}$) is
generally limited by the Eddingtom accretion rate
$\dot{M}_{\rm E}\simeq 10^{18} {\rm g\,s^{-1}}\simeq 10^{-8}\,M_\odot\, \rm yr^{-1}$.
The initial transfer rate of the fallback material to the central object
can be hypercritical and greatly exceed the Eddingtom limit.
For example, in the case of SN 1987A, \citet{che89} suggested that the transfer rate
at the time of reverse shock is $\sim 2.2\times 10^{28}\rm \,g\,s^{-1}$.
More recently, \citet{zwh08}
studied the supernova fallback for a wide range of progenitor masses and
various metallicities and explosion energies. The transfer rate was also found to be
$\sim 10^{29}\rm \,g\,s^{-1}$ at the early phase.
Note that this hypercritical accretion usually lasts very short time (less than 1 yr)
compared to the spin evolution time of NSs. Super-Eddinton accretion disks are likely
to be advective and emit a wind.
In the adiabatic inflow-outflow solutions (ADIOS)
the mass transfer rate varies radially as $\dot{M}(r)\sim \dot{M}_{\rm tr}(r/R_{\rm d})^p$
with $0\le p \le 1$  \citep{bb99}.
The case $p = 0$ corresponds to the absence of a wind, while $p = 1$ implies strong mass loss,
and the mass transfer rate at the inner edge $\dot{M}_{\rm in}$ is then not equal
to $\dot{M}_{\rm tr}$. A convenient form to relate the two rates is as follows,
\begin{equation}
\dot{M}_{\rm in} =
\left\{
\begin{array}{ll}
\dot{M}_{\rm E}(\dot{M}_{\rm tr}/\dot{M}_{\rm E})^s, \; & {\rm for}\
\dot{M}_{\rm tr} > \dot{M}_{\rm E},\\
\dot{M}_{\rm tr}, \; & {\rm for}\ \dot{M}_{\rm tr} \leqslant\dot{M}_{\rm E},
\end{array}
\right.
\end{equation}
with $0\le s \le1$. For example, if $s=0.75$, an initial $\dot{M}_{\rm in}$
distributed between $10^{25}$ gs$^{-1}$ and $10^{28}$ gs$^{-1}$ (as used below) corresponds to
a $\dot{M}_{0}$ between $2\times 10^{27}$ gs$^{-1}$ and $2\times 10^{31}$ gs$^{-1}$,
or the fallback mass between $4\times 10^{-6}\,M_{\sun}$ and $0.04\,M_{\sun}$
for typical disk formation time of about 1 s. Since the value of $s$ (or $p$) is highly uncertain
and it is unclear how much of the wind matter really leaves the system or falls back
again to the disk, we don't consider the winds from the fallback disk for simplicity
(i.e., $s=1$ or $p=0$),
except the mass loss due to the Eddington limit accretion.
In this case the accretion rate $\dot{M}_{\rm acc}$ of the NS is usually
assumed to be $\dot{M}_{\rm acc}=\min[\dot{M}_{\rm E},\dot{M}_{\rm in}]$.
However, if the photon's optical depth is too high, the
radiation is trapped within a radius at which the outward diffusion luminosity equals the
inward convected luminosity, so that the neutrino loss plays a crucial role, and the
radiative transfer probably only dominates in the later evolution \citep{col71,che89}.
\citet{che89} showed that when the mass transfer rate decreases to
$\dot{M}_{\rm cr}\simeq 3 \times 10^{-4}\,M_\odot\, \rm yr^{-1}$ the trapped photons
can escape from the shocked envelope and the Eddington limit is enabled. Thus
we assume that above this value the transferred mass is all accreted by the NS.
The accretion rate of the NS can be formulated as
\begin{equation}
\dot{M}_{\rm acc}=
\left\{
\begin{array}{ll}
\dot{M}_{\rm in}, \; & {\rm for}\
\dot{M}_{\rm in} > \dot{M}_{\rm cr} \; {\rm or} \; \dot{M}_{\rm in} < \dot{M}_{\rm E},\\
\dot{M}_{\rm E}, \; & {\rm for}\ \dot{M}_{\rm E} \leqslant \dot{M}_{\rm in} \leqslant \dot{M}_{\rm cr}.
\end{array}
\right.
\end{equation}

%Unlike the case in X-ray binary system, there is no donar star in
%NS and fallback disk system and the mass transfer rate at outer and inner
%annulus of the disk maybe different. Since the only things that matter is the }

\section{Spin and magnetic field evolution of NSs}
\subsection{Spin evolution of NSs with a fallback disk}
The spin evolution of NSs with a fallback disk can be divided into three phases according
to the relationship of the following three critical radii.

(1) The magnetospheric radius
$R_{\rm m}$, at which the ram pressure of the accretion flow equals
the magnetic pressure. Here we assume (Ghosh \& Lamb 1979)
\begin{equation}
R_{\rm m}=0.5R_{\rm A}=0.5(\frac{B^2R^6}{\dot{M}_{\rm in}\sqrt{2GM}})^{2/7}, \nonumber
\end{equation}
where $R_{\rm A}$ is the Alfv{\'e}n radius, and $G$ is the gravitational constant.
Note that if the mass transfer rate is sufficiently high $R_{\rm m}$ may be smaller than
the NS radius $R$. This is physically impossible thus in our calculation we use
\begin{equation}
R_{\rm m}={\rm max}(R,0.5R_{\rm A}) \nonumber
\end{equation}
as the value of magnetospheric radius.

(2) The corotation radius
\begin{equation}
 R_{\rm co}=(GM/\Omega^2)^{1/3}, \nonumber
\end{equation}
at which the Keplerian
angular velocity of the disk is equal to that of the NS.

(3) The light cylinder radius
\begin{equation}
R_{\rm LC}=c/\Omega, \nonumber
\end{equation}
at which the corotating extension of the NS is equal to the speed of light $c$.

We assumed that the pulsar activity is switched off and a disk torque
is exerted on the NS when the disk is able to penetrate into the light
cylinder (i.e., $R_{\rm m} < R_{\rm LC}$). Furthermore, if
$R_{\rm m} < R_{\rm co}$ the NS is in the accretor phase, otherwise the accretion
flow is stopped and ejected by the centrifugal barrier and the NS is in the propeller
phase \citep{is75}. The (unified) disk torque exerted on the NS in these two cases is taken to be
\citep{mph01},
\begin{equation}
I\dot{\Omega}=2\dot{M}_{\rm acc}R_{\rm m}^2\Omega_{\rm K}(R_{\rm m})\left[1-\Omega/\Omega_{\rm K}(R_{\rm m}) \right],
\end{equation}
where $\Omega_{\rm K}$ is the Keplerian angular velocity
in the disk. The NS is spun-up when $\Omega < \Omega_{\rm K}(R_{\rm m})$ and
spun-down when $\Omega > \Omega_{\rm K}(R_{\rm m})$. When $\Omega = \Omega_{\rm K}(R_{\rm m})$
the NS is spinning at the so-called ``equilibrium period'' given by
\begin{equation}
P_{\rm eq}=2\pi \left(\frac{R_{\rm m}^{3}}{GM}\right)^{1/2}.
\end{equation}

If $R_{\rm m} \geqslant R_{\rm LC}$ the pulsar activity starts to work,
and the NS is in the ejector phase.
Since the kinetic energy density in the disk has
a radial dependence $\propto r^{-5/2}$ (where $r$ is the distance from the center
of the NS) steeper than the electromagnetic energy density
outside the light cylinder ($\propto r^{-2}$), stable
equilibrium of the disk outside the light cylinder is not allowed,
unless it is beyond the gravitational capture radius \citep{lbw92}.
So in this case we consider the spin-down torque only due to magnetic dipole
radiation\footnote{This is different from \citet{mph01} and \citet{yps12},
who assumed that even in the ejector phase the disk can still exist outside the
light cylinder, and
the inner radius of the disk always coincides with the light cylinder radius
irrespective of the decreasing mass transfer rate.}.

As the pulsar ages the radio luminosity will fade away, and finally it will
become unobservable as a pulsating source. This is assumed to occur when its evolutionary
track on the $B-P$ plane crosses the so-called death-line given by \citep{rs75}
\begin{equation}
B =0.17\times 10^{12}P^2\,{\rm G}.
\end{equation}

\subsection{Magnetic field evolution}
Magnetic field is the most important parameter that
determines the spin evolution and the observational properties of isolated NSs.
We assume that, along with accretion, the NS magnetic fields decay with the following
form \citep{tv86,sms89}:
\begin{equation}
B=\frac{B_0}{1+\Delta M/10^{-5}M_{\odot}},
\end{equation}
where $B_0$ is the initial magnetic field and $\Delta M$ is the accreted mass.
When accretion stops the buried field will re-diffuse to the surface due to Ohmic
diffusion and Hall drift \citep[e.g.,][]{gpz99}. The field diffusion process is
governed by the MHD induction equation, and its speed depends on
the initial magnetic strength
and the overall accreted mass \citep[for recent reviews, see][]{gep09,ho13}.
Based on the numerical calculations of field re-diffusion
\citep[e.g.][]{gpz99,ho11}, we fit the results with a phenomenological law
for the growth rate of the magnetic field after accretion,
\begin{equation}
\dot{B}=0.01\left( \frac{\Delta M}{M_{\odot}} \right)^{-3} \left(1-\frac{B}{B_0}\right)^2\;{\rm G\,yr^{-1}}.
\end{equation}
Figure 1 illustrates the field emergence with different accreted mass.

\section{Synthesis of NS population}
In our Monte-Carlo simulation of the NS population we adopt similar input
parameters as in \citet{yps12}. In previously
works, the distribution of the birth spin periods  of NSs has been suggested to
be in a wide range from several milliseconds to hundreds of milliseconds
\citep[e.g.][]{acc02,f-gk06}, with various forms, e.g., a fix value
\citep[e.g.][]{fcm01}, a flat distribution \citep[e.g.][]{khb08}, and a
(logrithm) Gaussian distribution \citep[e.g.][]{rdfp01,lbd93}.
The SNRs may present useful constraints on the initial
spins of the pulsars associated with them.
There is evidence supporting that the NSs are born with both fast and slow rotation.
For example, the initial spin period ($\sim 19\rm \, ms$) of the Crab pulsar
estimated from the age of the Crab nebula and the measured braking index
\citep{mt77} implies that the NS was born spinning rapidly.
The CCO 1E 1207.4$-$5209 in the SNR
G296.5+10.0 has a spin period of 0.424 s \citep{zps00} and period derivative
$\dot{P} < 2.5 \times 10^{-16}$ ss$^{-1}$ \citep{gh07}.
The characteristic age of the NS $\tau_{\rm c} >27$ Myr exceeding
the age of the SNR by three orders of magnitude, suggesting that
1E 1207.4$-$5209 was born with a spin period very similar to the current value.
Here we adopt three different distributions of the initial spin periods,
i.e., the ``fast" population with,
\begin{align}
\langle{\rm log}P_0({\rm s})\rangle=-2.3,\;\sigma_{{\rm log}P_0}=0.3,
\end{align}
the ``slow" spin population with
\begin{equation}
\langle{\rm log}P_0({\rm s})\rangle=-0.5,\;\sigma_{{\rm log}P_0}=0.2,
\end{equation}
and the ``composite'' population in which we assume that $\sim 40\%$ of the pulsars
are born in the slow population \citep{vml04}.
%\textbf{So in
%``composite'' population there are only pulsars, that is to say all
%the NSs must evolve to ejector phase.}

For the initial magnetic field almost all the works adopted a logarithm
Gaussian distribution, and we take the following form of \citet{acc02}:
\begin{equation}
\langle{\rm log}B_0({\rm G})\rangle=12.35,\;\sigma_{{\rm log}B_0}=0.4.
\end{equation}

We follow \citet{yps12} to assume a logarithm uniform distribution of the initial
mass transfer rate $\dot{M}_0$ ranging from $10^{25}$ to $10^{28}$ $\rm g\,s^{-1}$, which
is roughly consistent with previous semi-analytical and numerical results
\citep{che89,mwh01,mnh97,zwh08}.
The initial radius of the fallback disk depends on the specific angular momentum
of the fallback material. At the pre-supernova stage the specific angular momentum
of the iron core of a rapidly rotating star with mass $8-25M_\odot$ is
$\sim 10^{16}-10^{17}\rm cm^2\,s^{-1}$ \citep[see][]{hlw00}, corresponding a
circularization radius $\sim 10^6-10^8\rm \, cm$. Thus we randomly  select the logarithm
of the initial radius (in units of cm) between 6 and 8.

Figures 2-4 illustrate the evolution of NSs with different initial parameters. Here we
take typical values for the mass ($1.4M_\odot$), radius ($10^6\,\rm cm$) and
initial magnetic field ($\sim 2\times 10^{12}$ G) of the NS.
The braking index $n$ and characteristic age $\tau$ are plotted
throughout the evolution although they are measurable only
in the ejector phase. We consider three cases, in two of which
$\dot{M}_0$ and $R_0$ are close to the maximum and minimum of the adopted values
(note that $t_{\rm v}$ and $t_{\rm d}$ are positively correlated with $R_0$),
and in third one we choose the medium values for $\dot{M}_0$ and $R_0$.
In each figure the initial spin period is taken to be $P_0 = 300$ ms and 5 ms
in the left and right panels, and the accretor, propeller and ejector phases are
shown in dotted, dashed and solid lines, respectively.

In Fig.~2 we set $\dot{M}_0=10^{28}\,\rm gs^{-1}$ and $R_0=10^8\,\rm cm$
for NSs with both fast and slow initial spins. The magnetic field decays to
be below $10^{9}\,\rm G$
within $\sim 0.1$ yr as $\sim 4\times 10^{-2}M_\odot$ mass is accreted,
and has not recovered to its initial value at $10^6$ yr.  In the left panel
the  NS is correspondingly accelerated from 300 ms to $\sim 4$ ms within the first $\sim 0.1$ yr.
The accretor phase lasts $\sim 10^5$ yr, followed by the propeller phase until $10^6$ yr.
The time dependence of $n$ and $\tau_{\rm c}$ is more complicated, which vary
drastically during the evolution. At the beginning of the evolution,
due to the enormously high accretion rate, the magnetosphere radius is suppressed
to the NS surface. In this early stage $n<0$ since $\ddot{\Omega}<0$.
When $\dot{M}_{\rm in}$ decreases
to be lower than $\dot{M}_{\rm cr}$, $\dot{M}_{\rm acc}$ is limited by the Eddington limit.
At this time $R_{\rm m}$ is still equal to $R$ (note that $R_{\rm m}$ depends on
$\dot{M}_{\rm in}$), and $\dot{\Omega}$ is nearly constant, thus $n\sim 0$.
As $\dot{M}_{\rm in}$ continues to decrease $R_{\rm m}$  becomes larger than $R$,
and $n$ rapidly rises to $\sim 50$ because $\dot{R}_{\rm m}>0$
and $\ddot{M}_{\rm acc}\sim 0$ at this time.
When $\dot{M}_{\rm in}=\dot{M}_{\rm E}$, $n$ has another abrupt change.
At this time $\ddot{M}_{\rm acc}$ dominates over $\dot{R}_{\rm m}$ and $n<0$.
When $\dot{M}_{\rm acc}$ decreases so that $R_{\rm m}= R_{\rm co}$
the NS enters the propeller phase, and the evolution is stopped at $10^6$ year.

In Fig.~3 we set moderate values for $\dot{M}_0(=10^{27}\,\rm g\,s^{-1})$ and
$R_0(=10^7\,\rm cm)$. Since the accreted mass ($\sim 10^{-3}M_\odot$) is
significantly lower than in Fig.~2,
the magnetic field first decays to a few $10^{10}$ G, and re-grows to its initial value
within $10^6$ yr. The fast-spinning NS has experienced all three evolutionary phases
while the slow one is still in the propeller phase at the age of $10^{6}$ yr.
In Fig.~4 we take $\dot{M}_0=10^{25}\rm\,g\,s^{-1}$  and
$R_0=2 \times 10^6\rm\,cm$  (we note that  if $R_0=10^6\rm\,cm$, $R_{\rm m}$ is
larger than $R_0$ at the beginning of the evolution, the fallback disk does not exist
and the NS enters the ejector phase directly). With such a low mass transfer rate
the accretor phase is very short, about 2 yr and $10^{-3}$ yr for slow and rapid spinning
NSs, respectively. The ejector phase correspondingly starts at around $10^4$ yr and 0.1 yr.

As we are only concerned with young pulsars, in the Monte-Carlo simulation
we generate 100 NSs every 100 years in the first $10^4$ years, and extrapolate the
results to $10^6$ years by sampling the parameters of NSs every $10^4$ years
during the calculation. The distributions of $P$ and $B$ of the NS population
are plotted in Fig.~5, at the time when the NS enters the pulsar phase,
and of $10^4$ yr and $10^6$ yr. Note that here ``pulsars" mean NSs in the
ejector phase and
located above the death line in the $B-P$ diagram.
The calculated $P$ distribution at $10^6$ years shows double-peak
structure for both the fast and slow spin populations. The reason is as follows.
When the initial mass transfer rate is high,
the NS magnetic field is decayed by several orders of magnitude, and
the NS spends most of the time (several $10^5$ yr) in the accretor
and propeller phases. Because of the weak magnetic field,
the spin-down induced by the propeller torque and magnetic dipole radiation
is insignificant, giving rise to
a group of rapidly spinning pulsars in both the fast and slow population.
The distribution of the composite population is a hybrid of the
fast and slow population and has
multi-peak structure at $10^6$ yr.
%There is no death pulsars in
%our results since those ``dead pulsars'' are still in accretor or propeller phase.
For the fast spin population the distributions of $P$ and $B$ at $10^4$ yr
provide a natural transition between the initial pulsar state
to the relatively old one (at $10^6$ yr).
However it is not the case in the slow spin population because,
unlike the fast spin population in which
most NSs have evolved into the ejector phase at $10^4$ yr, many of the slow NSs
are still in the accretor and propeller phases.

In order to show the distribution of the braking indices and the
characteristic ages for pulsars with different ages, we adopt the same
initial distribution of the parameters as in the previous simulations but
a variable birth rate, to ensure that the total number of NSs is constant for every
logarithm interval of the age \citep[see also][]{yps12}. For example,
for ages between $10-10^{2}\,\rm yr$, $10^{2}-10^{3}\,\rm yr$ and
$10^{3}-10^{4}\,\rm yr$ the birth rate is taken to be $1\,\rm yr^{-1}$,
$10^{-1}\,\rm yr^{-1}$ and
$10^{-2}\,\rm yr^{-1}$ respectively.
Figure 6 shows the distributions of $\tau_{\rm c}$ and $n$ versus the age.
The green and red crosses represent the NSs as observable pulsars and in the
accretor/propeller phase, respectively. It is seen that for the slow spin population
very few NSs can enter the pulsar phase within $10^6$ years.

For comparison with observations we also plot the observational data of pulsars
with dots (with error bars) in Fig.~6. Here the pulsars are those associated with
SNRs and with measured braking indices. The samples
of SNR-PSR associations are taken from the ATNF pulsar catalogue\footnote{http://www.atnf.csiro.au/people/pulsar/psrcat/}
and a census of high-energy observations of the Galactic SNRs\footnote{http://www.physics.umanitoba.ca/snr/SNRcat/} \citep{ATNF,HEcat-GSNR}.
Thus we use the SNR ages as the real ages of the pulsars.
The parameters of these pulsars are listed in Table 1.
%We compare the samples with the calculated spin period distribution of
%NSs when they become pulsars, our results shows that the distribution of
%$P_0$ of slow spin population is more compatible with the samples of SNR-PSR association.
%The NSs with measured SNR ages and braking indices are also illustrated in figure 6.
%The observed distribution of $\tau$ and $n$ seems not compatible with the slow spin population.

%For slow spin population both the distributions of $\tau$ and $n$ are more dispersed,
%since for slow spin population the majority of the NSs are in accretor or propeller phase and
%the time dependence of $\dot{P}$ and $\ddot{P}$ is more complicated than in ejector phase.
For all the three populations, the characteristic ages of the pulsars usually exceed their
real ages  by a factor up to $\sim 10^3$.
The braking indices concentrate in the range $-1<n<3$ which is consistent
with observations, but they can reach $\sim -10^3$.
There is no pulsar with $n>3$, since in our model NSs with $n>3$ are
in the accretor or propeller phase and thus unobservable, and the magnetic field
growth is the only cause for the variation of $n$.
Actually the braking
indices for NSs in accretor and propeller phase are distributed in a much wider range
of $[-10^5,10^3]$.

\section{Discussion and conclusions}

We have carried out population synthesis calculations of the NS evolution,
taking into account the supernova fallback accretion, which suppresses the NS magnetic fields,
and the post-emergence of the buried field.
Though the input parameters in our simulation are similar to those in \citet{yps12},
some of the important assumptions are different in the two works.
First, \citet{yps12} assume that a fallback disk always surrounds the NS and
exerts a torque on it; in our model the evolutionary sequence of the NS is divided into
the accretor,
propeller and ejector phases, the disk torque works only in the
accretor and propeller phases, and the NS acts as a pulsar only in the ejector phase.
Second, we consider the hypercritical accretion from the fallback matter
when $\dot{M}_{\rm in}>\dot{M}_{\rm cr}$,
and include the magnetic field evolution during and after the accretion,
while in \citet{yps12} the mass accretion rate is limited by $\dot{M}_{\rm E}$ and
the magnetic field is assumed to be constant. Third, for the disk evolution we take the formation
time $t_0 = {\rm max}(t_{\rm v},t_{\rm d})$ rather
$t_0 = t_{\rm d}$ as in \citet{yps12}. This results in a lower mass transfer rate
in the disk at the same age in our case.
In our work due to the magnetic field re-emergence $n$ is always smaller than 3
when the NS is in the ejector phase, while in \citet{yps12} the deviation of $n$
from 3 is caused by the
disk torque: the majority of the pulsars have $n<3$, but a considerable fraction
of them have $n>3$ (especially for the slow spin population).
The distributions of both $\tau_{\rm c}$ and $n$ in our work are more dispersed and extended
than in \citet{yps12}, since the magnetic field usually evolves on a longer timescale than
the fallback disk.

Figure 6 shows that the statistical results of the fast spin population seem to be
compatible with the observed distribution of the braking indices and the ages of
young pulsars. NSs born with slow spins are difficult
to survive the accretor and propeller phases. On the other hand, the birth period
distribution for pulsars may also present possible constraints on the
initial period distribution of the NSs\footnote{Note that the birth periods
of pulsars are the initial periods when the NSs enter the ejector phase. They are
similar to but
not identical with the initial periods of newborn NSs.}. Recently
\citet{nsk13} derived the kinematic ages for 52 pulsars based on the
measured pulsar proper motions and positions, by modelling the trajectory of the
pulsars in a Galactic potential. They found that the birth periods of these pulsars
show two-population structure, one with $P_0<400$ ms and
the other with 700 ms $<P_0<1.1$ s. Although this result is based on the
standard magnetic-dipole braking ($n = 3$), it is unlikely to deviate far from
the real situation as shown by the authors. If we compare their birth period distribution
with our calculated results (Fig.~5), we find that
a considerable fraction of NSs may be born with relatively slow spins. This suggests
that the newborn NSs might be composed by composite populations with a wide range
of the spin periods.

It should be noted that currently solid theories on SN fallback and related field
evolution are lacking,
and hence many parameters adopted in our model are quite uncertain.
This means that our results can be only regarded as illustrative rather for
real situation.
However, they keep the basic features for the evolution of the braking indices and the relation
between the characteristic ages and the real ages. The model may be \textbf{tested} or refined
by future observations.

We do not consider late evolution of the NSs. As shown by
\citet{pvg12},
the effect of the magnetic field evolution on the braking index can be divided into
three qualitatively different stages, depending on the age and the internal temperature
of the NS:
a first stage with fallback accretion and subsequent field evolution
($n < 3$); in a second stage, the evolution is governed by almost pure Ohmic field decay,
and a braking index $n > 3$ is expected; in the third stage, at late times, when the interior
temperature has dropped to very low values, the Hall oscillatory modes in the NS
crust result in braking indices of high absolute value and both positive and negative signs.
In the model proposed by \citet{zx12} the field evolution is caused by a long-term power law
decay coupled with short-term oscillations.

\begin{acknowledgements}

This work was supported by the Natural Science Foundation of China
under grant numbers 11133001 and 11203009, the
National Basic Research Program of China (973 Program 2009CB824800),
and the Qinglan project of Jiangsu Province.

\end{acknowledgements}

%\bibliographystyle{apj}
%\bibliography{my}

\clearpage

\begin{figure}
\centering
\includegraphics[width=13cm]{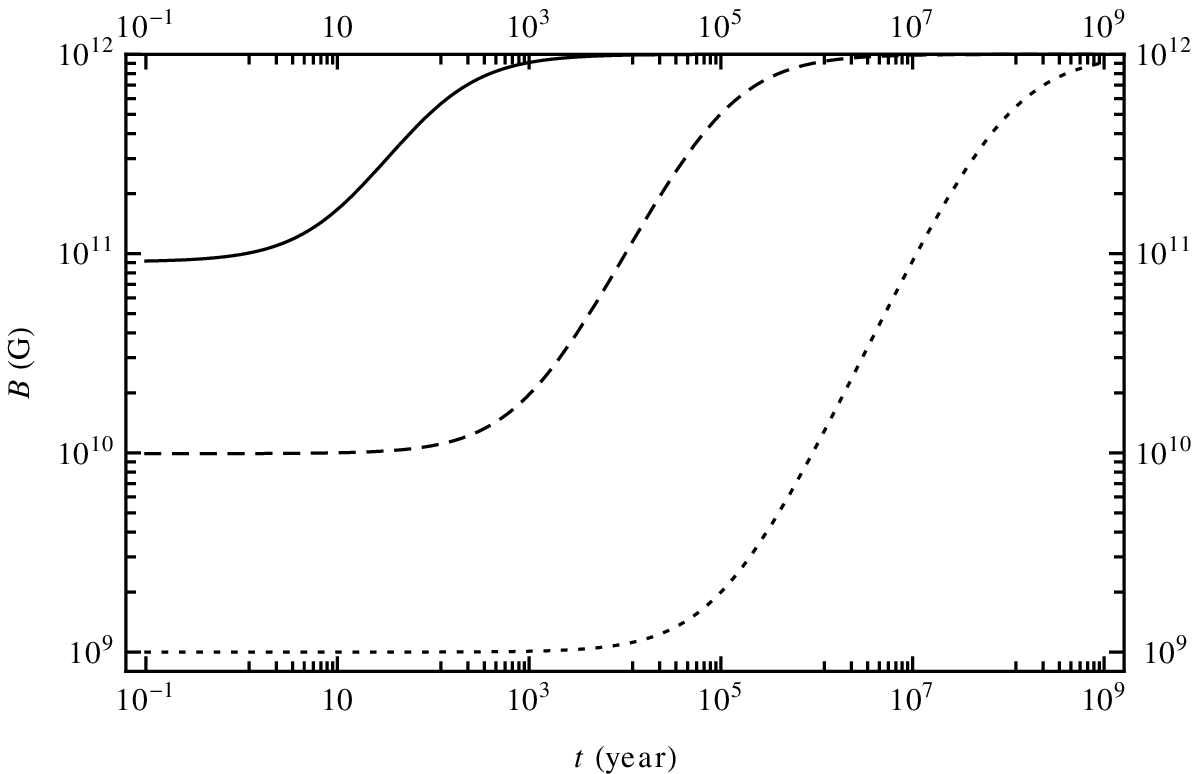}
\caption{An illustration of the magnetic field emergence according to
Eq.~(15) for an initial magnetic
field strength of $B_0=10^{12}\,\rm G$. The lines from top to bottom
correspond to the accreted mass of $10^{-4}$,
$10^{-3}$ and $10^{-2}$ $M_\odot$, respectively.}
\end{figure}
\begin{figure*}
\centering
\includegraphics[width=8.1cm]{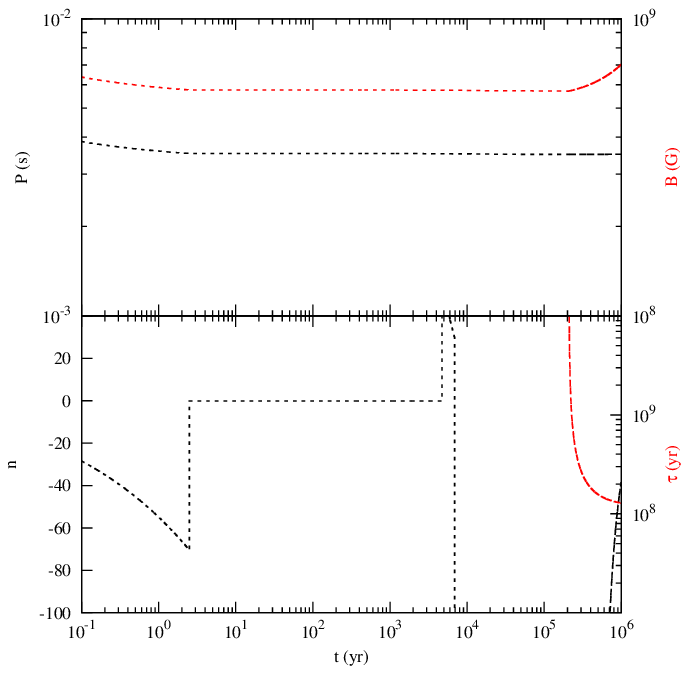}
\includegraphics[width=8.1cm]{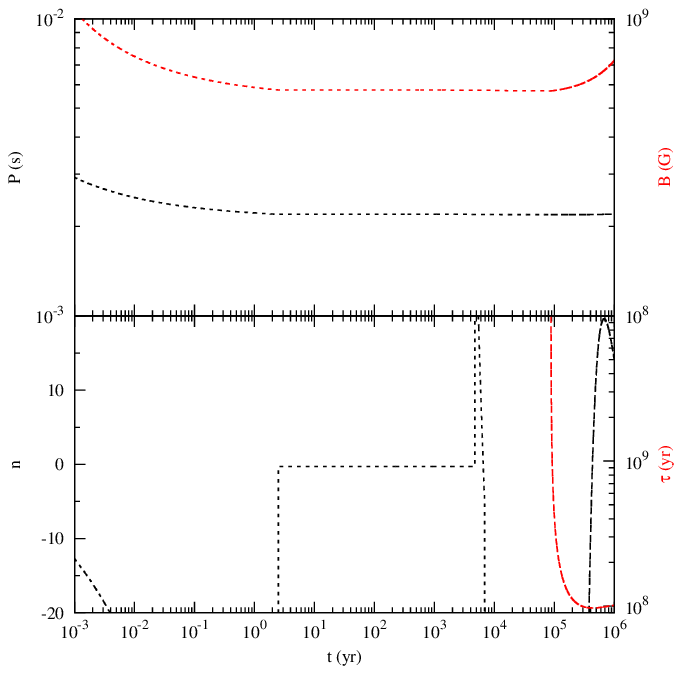}
\caption{Evolution of the spin period $P$, the
magnetic field $B$ (top), the braking index $n$ and the characteristic age
$\tau$ (bottom) of a NS with an initial magnetic field of
$2.24 \times 10^{12}\,\rm G$.
The initial spin period is taken to be $P_0=300\,\rm ms$ and
$5\,\rm ms$ in the left and right panels, respectively; in both panels
we set $R_0=10^8\,\rm cm$ and $\dot{M}_0=10^{28} \rm g\,s^{-1}$.
%The accretor, propeller and
%ejector phases are represented with dotted, dashed and solid lines, respectively.
In this and the following two figures,
the curves for $P$ and $n$ are plotted in black, and for $B$ and $\tau$ in red;
the accretor, propeller and ejector phases are represented with dotted,
dashed and solid lines, respectively.
}
\end{figure*}
\clearpage

\begin{figure*}
\centering
\includegraphics[width=8.1cm]{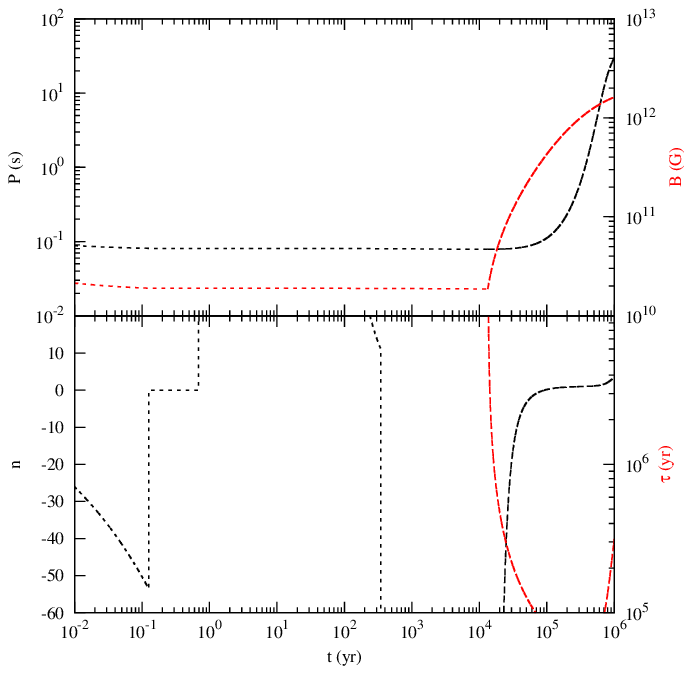}
\includegraphics[width=8.1cm]{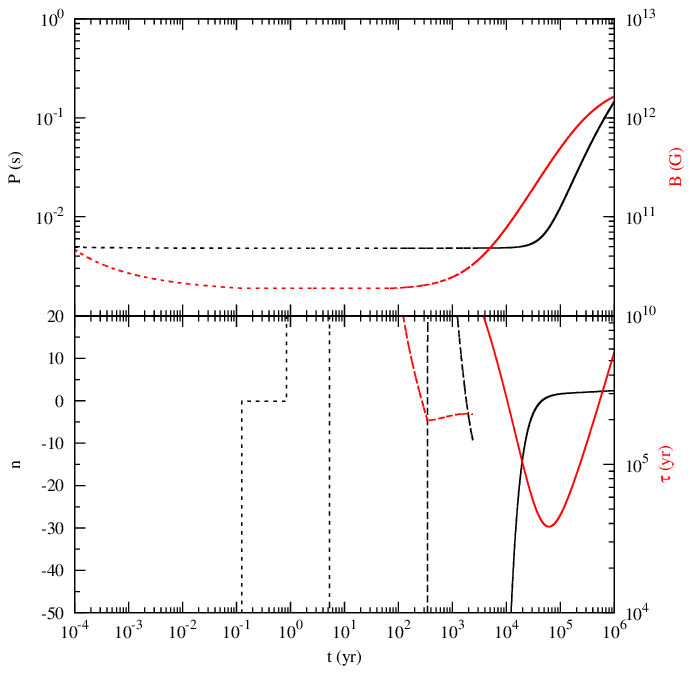}
\caption{Same as Fig.~2 but for
$R_0=10^7 \rm cm$ and $\dot{M}_0 = 10^{27} \rm g\,s^{-1}$.}
\end{figure*}
\clearpage

\begin{figure*}
\centering
\includegraphics[width=8.1cm]{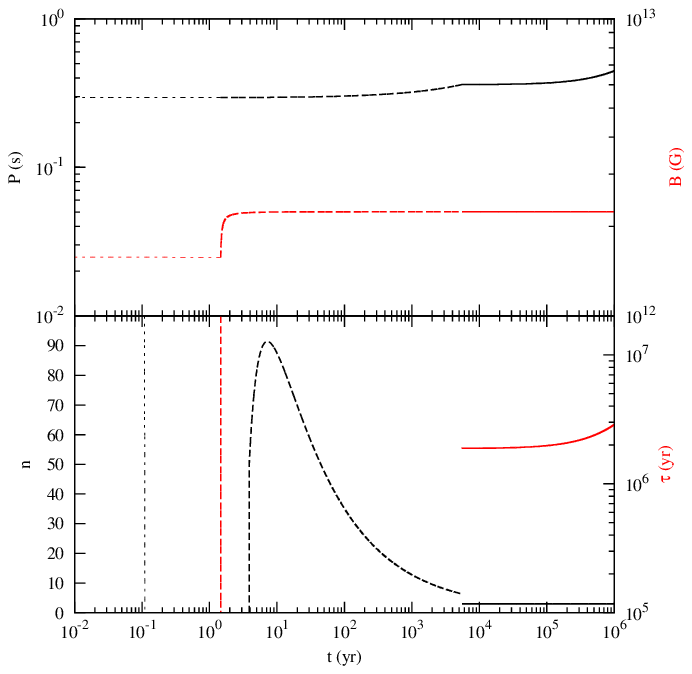}
\includegraphics[width=8.1cm]{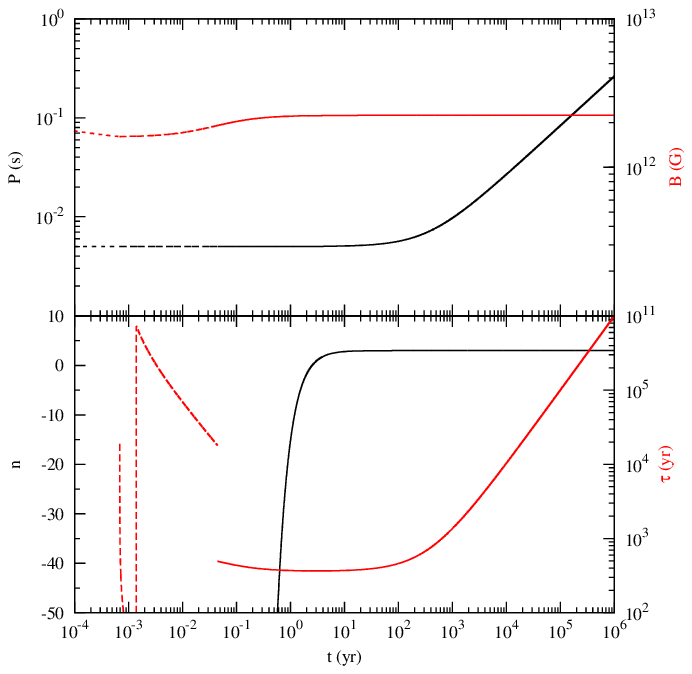}
\caption{Same as Fig.~2 but for
$R_0=2\times10^6 \rm cm$ and $\dot{M}_0 = 10^{25} \rm g\,s^{-1}$.
}
\end{figure*}
\clearpage

\begin{figure*}
\centering
\includegraphics[width=7.7cm]{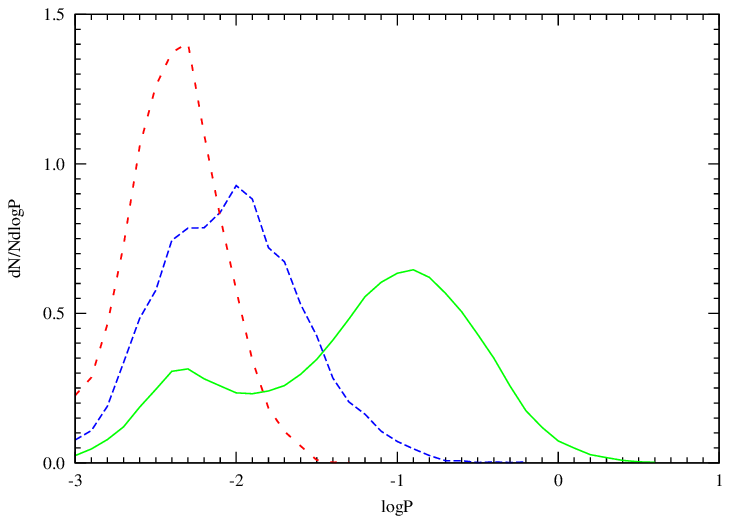}
\includegraphics[width=7.7cm]{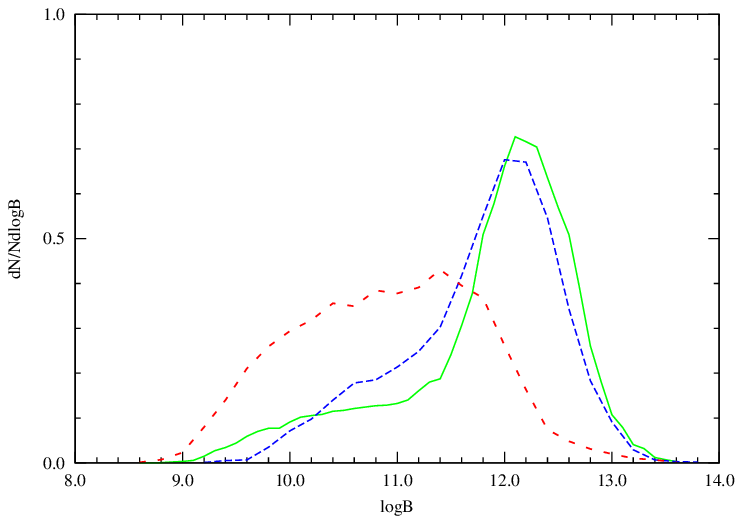}\\
\includegraphics[width=7.7cm]{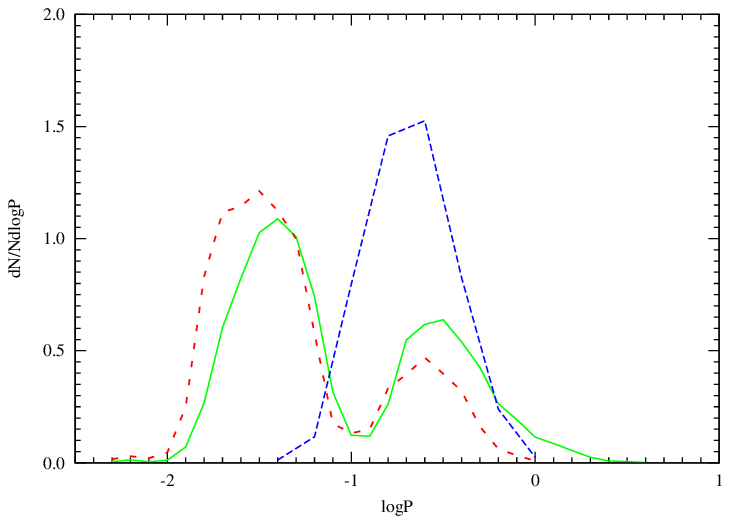}
\includegraphics[width=7.7cm]{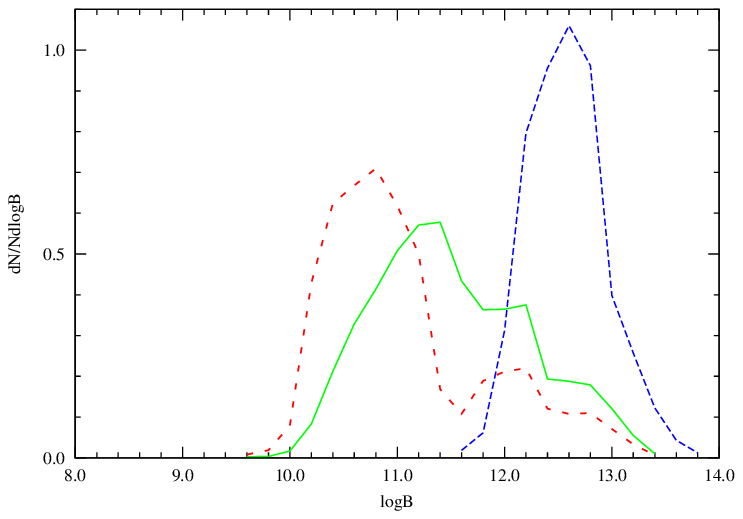}\\
\includegraphics[width=7.7cm]{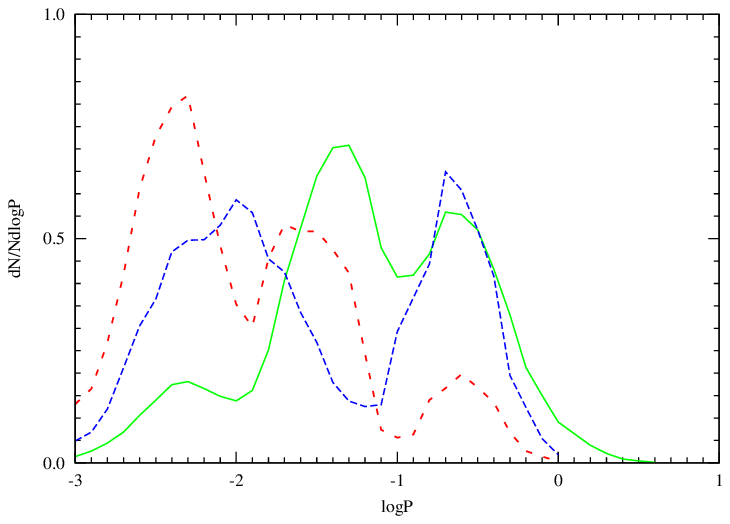}
\includegraphics[width=7.7cm]{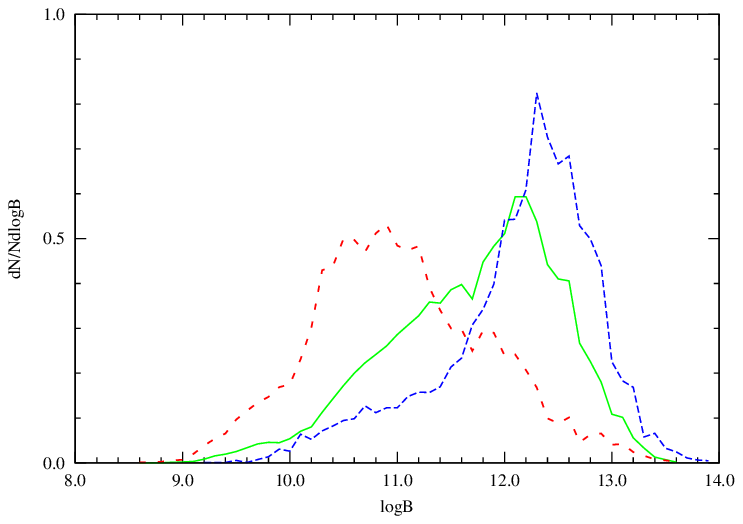}
\caption{The distribution of the spin period $P$ and the magnetic field $B$ of the pulsars.
From top to bottom panels are the fast, slow and composite spin populations, respectively.
%The NSs in ejector phase is plot in green and which in accretor or propeller phase
%is unobservable to us and plot in red.
Dotted lines in red represent the distributions of the initial spin period and
magnetic field of pulsars, i.e.,
when the NSs enter the ejector phase.
Blue dashed and green solid lines are for the distributions of $P$ and $B$ of pulsars
at $10^4$ and $10^6$ yr.
The composite population is composed of 60$\%$
fast spulsars and 40$\%$ slow pulsars.}
%Histogram in violet is plot for the NSs associated with SNRs, for which $P_0$ is calculated
%from the measured braking index(for those $n$ are unavailable $n=2.5$ is assumed) and SNR
%age by using equation 2.}
\end{figure*}
\clearpage

\begin{figure*}
\centering
\includegraphics[width=7.0cm]{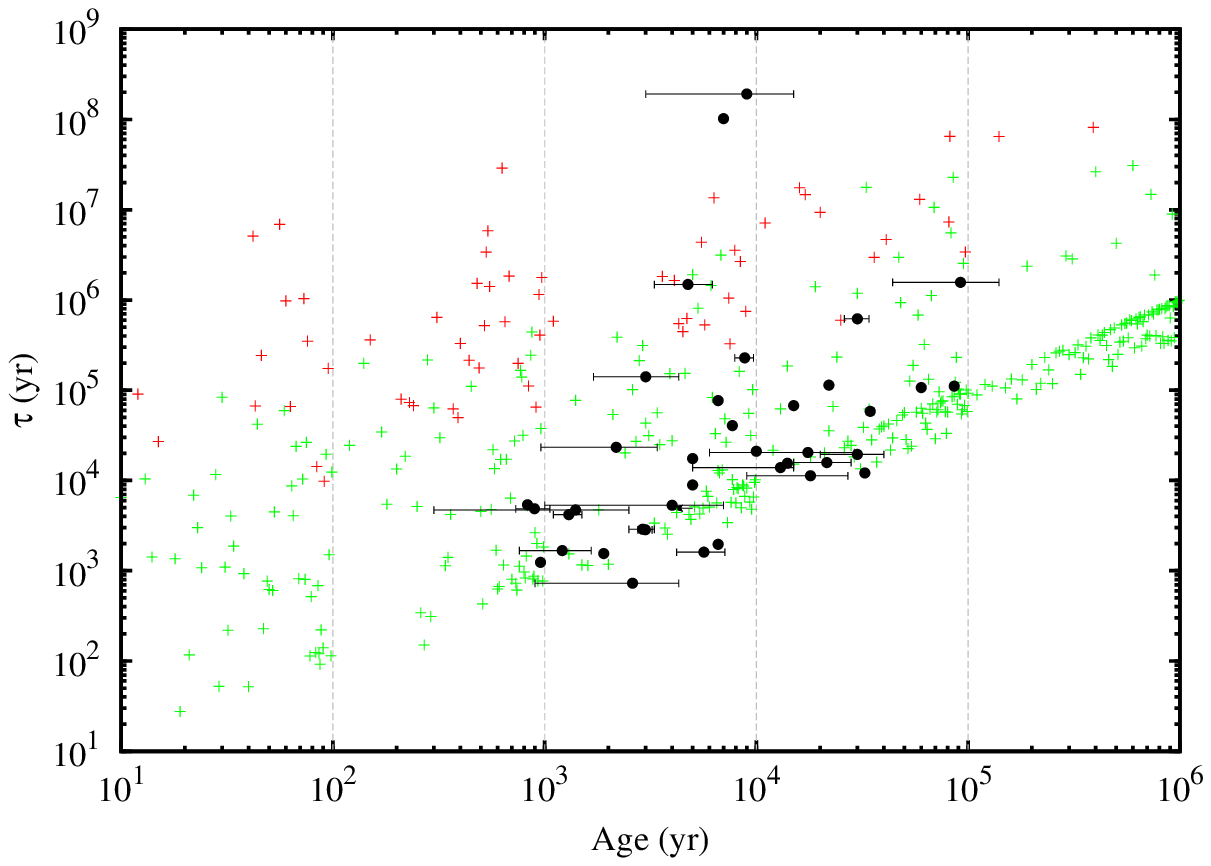}
\includegraphics[width=7.0cm]{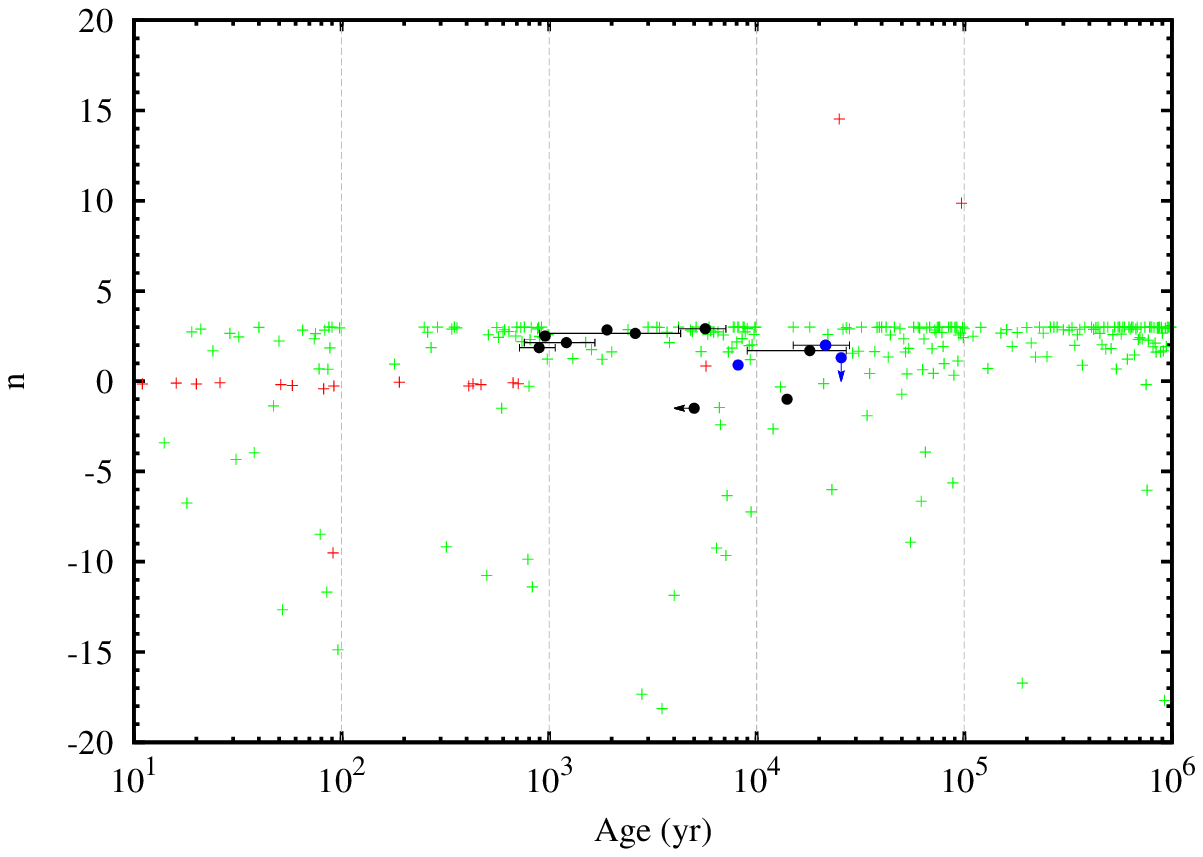}\\
\includegraphics[width=7.0cm]{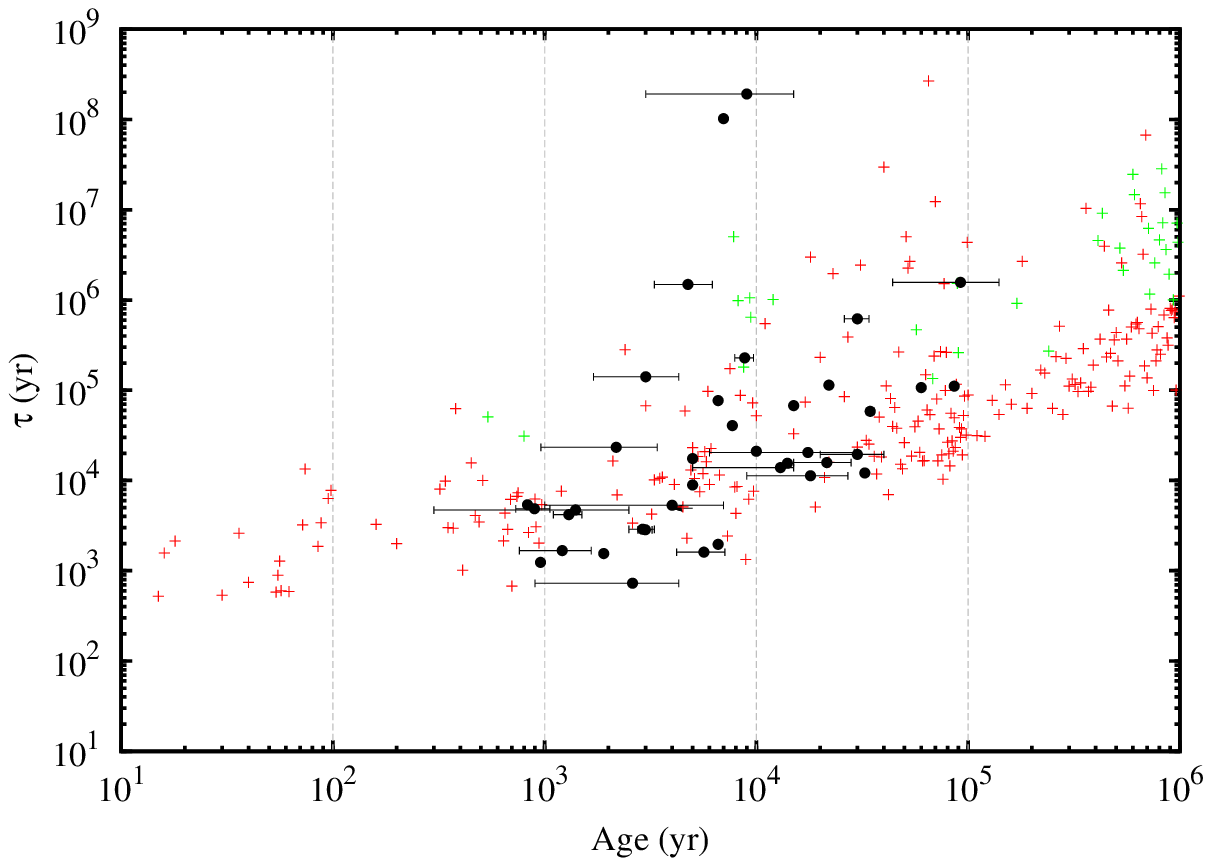}
\includegraphics[width=7.0cm]{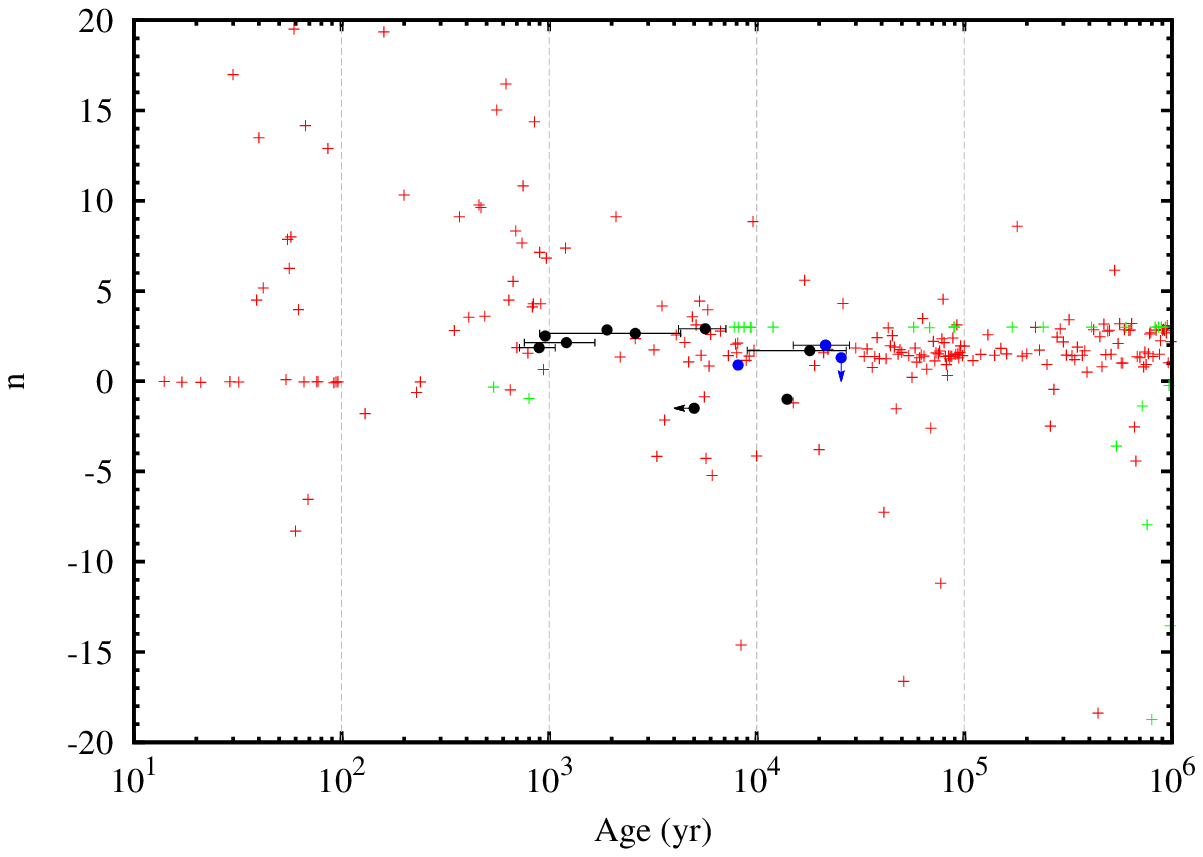}\\
\includegraphics[width=7.0cm]{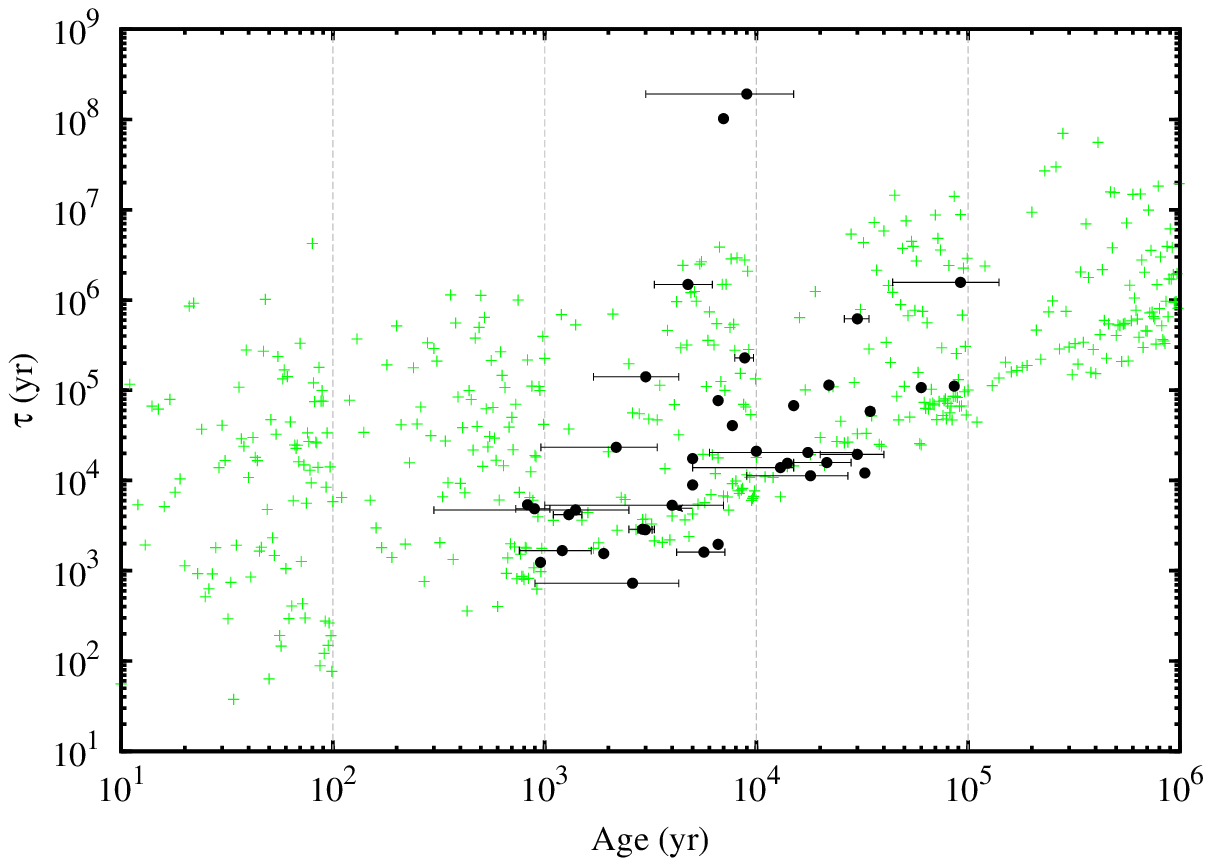}
\includegraphics[width=7.0cm]{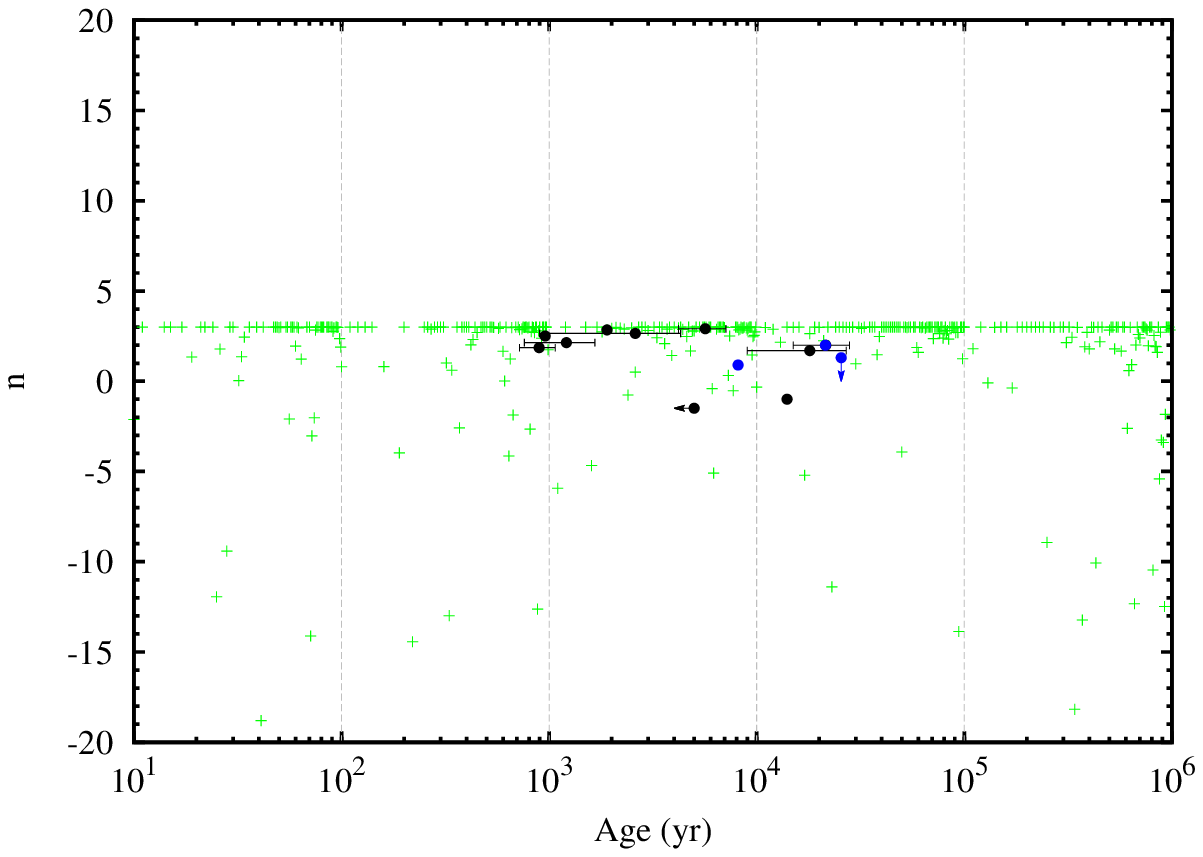}
\caption{A comparison of the distribution of $\tau$ and $n$ versus different NS ages.
From the top to the bottom panels are the fast, slow and composite spin populations.
NSs in the ejector phase are plotted in green crosses, those in the accretor
and propeller phases are plotted in red crosses.
NSs with measured braking indices and SNR associations are plotted in black
dots with error bars. For NSs associated with SNRs we use the SNR age as the real age
of the NS if the SNR age is available, except
for PSRs J1734$-$3333, J1747$-$2958 and B1823$-$13, for which
we use $\tau$ as the real age and plot with blue dots in the right panel.
The composite spin population is composed of 60$\%$
fast pulsars and 40$\%$ slow pulsars.}
\end{figure*}

\clearpage

\begin{deluxetable}{llcccccc}
\tabletypesize{\scriptsize}
\tablecolumns{8}
\tablewidth{0pt}
\tablecaption{Parameters for pulsars with measured braking indices and/or associated with SNRs}
\tablehead{
\colhead{PSR Name} & \colhead{$P$ (s)} & \colhead{$\dot{P}$} & \colhead{Assoc. SNR}
& \colhead{$\tau_{\rm c}\,{\rm (yr)}$} & \colhead{SNR age (kyr)} & \colhead{$n$} & \colhead{Ref.}\\
}
\startdata
J0007+7303 & 0.316 & 3.61E-13 & CTA1 & 1.39E+4 & $13^{+2}_{-8}$ & \nodata & 1 \\
J0205+6449 & 0.066 & 1.94E-13 & 3C58 & 5.37E+3 & 0.830,$\leqslant7$ & \nodata & 2\\
J0525-6607 & 8.047 & 6.50E-11 & N49 & 1.96E+3 & 6.6 & \nodata & 3\\
B0531+21 & 0.033 & 4.23E-13 & Crab PWN & 1.24E+3 & 0.957 & 2.51(1) & 4\\
J0537-6910 & 0.016 & 5.18E-14 & N157B,EXGAL:LMC & 4.93E+3 & $<5$ & -1.5 & 5\\
J0538+2817 & 0.143 & 3.67E-15 & S147 & 6.18E+5 & 26-34 & \nodata & 6\\
B0540-69 & 0.050 & 4.79E-13 & 0540-693,EXGAL:LMC & 1.67E+3 & $0.76-1.66$ & 2.140(9) & 7\\
J0821-4300 & 0.113 & 1.20E-15 & PUPPIS A & 1.49E+6 & 3.3-6.2 & \nodata & 8\\
B0833-45 & 0.089 & 1.25E-13 & Vela & 1.13E+4 & 9-27 & 1.7 & 9\\
J1016-5857 & 0.107 & 8.08E-14 & G284.3-1.8 & 2.1E+4 & 10 & \nodata & 10\\
J1119-6127 & 0.408 & 4.02E-12 & G292.2-0.5 & 1.61E+3 & 4.2-7.1 & 2.91(5) & 11\\
J1124-5916 & 0.135 & 7.53E-13 & G292.0+1.8 & 2.85E+3 & 2.93-3.05 & \nodata & 12\\
J1210-5226 & 0.424 & 6.60E-17 & G296.5+10.0 & 1.02E+8 & 7 & \nodata & 13\\
B1338-62 & 0.193 & 2.53E-13 & G308.8-0.1 & 1.21E+4 & 32.5 & \nodata & 14\\
J1437-5959 & 0.062 & 8.59E-15 & G315.9-0.0 & 1.14E+5 & 22 & \nodata & 15\\
B1509-58 & 0.151 & 1.54E-12 & G320.4-1.2(MSH 15-52) & 1.55E+3 & 1.9 & 2.839(1) & 16\\
J1550-5418 & 2.070 & 2.32E-11 & G327.24-0.13 & 1.41E+3 & \nodata & \nodata & \\
J1632-4818 & 0.813 & 6.50E-13 & G336.1-0.2 & 1.98E+4 & \nodata & \nodata & \\
J1635-4735 & 2.595 & \nodata & G337.0-0.1 & \nodata & \nodata & \nodata & \\
B1643-43 & 0.232 & 1.13E-13 & G341.2+0.9 & 3.25E+4 & \nodata & \nodata & \\
B1706-44 & 0.102 & 9.30E-14 & G343.1-2.3 & 1.75E+4 & 5 & \nodata & 17\\
J1726-3530 & 1.110 & 1.22E-12 & G352.2-0.1 & 1.45E+4 & \nodata & \nodata & \\
J1734-3333 & 1.169 & 2.28E-12 & G354.8-0.8 & 8.13E+3 & \nodata & 0.9(2) & 18\\
J1747-2809 & 0.052 & 1.56E-13 & G0.9+0.1 & 5.31E+3 & 1-7 & \nodata & 19\\
J1747-2958 & 0.099 & 6.13E-14 & PWN:G359.23-0.82 & 2.55E+4 & \nodata & $<$1.3 & 20\\
B1757-24 & 0.125 & 1.28E-13 & G5.4-1.2 & 1.55E+4 & 14 & -1 & 21\\
B1758-23 & 0.416 & 1.13E-13 & W28 & 5.83E+4 & 33-36 & \nodata & 22\\
B1800-21 & 0.134 & 1.34E-13 & G8.7-0.1 & 1.58E+4 & 15-28 & 2 & 23\\
J1808-2024 & 7.556 & 5.49E-10 & G10.0-0.3(W31) & 218 & \nodata & \nodata & \\
J1809-2332 & 0.147 & 3.44E-14 & G7.5-1.7 & 6.76E+4 & $\leqslant15$ & \nodata & 24\\
J1811-1925 & 0.065 & 4.40E-14 & G11.2-0.3 & 2.33E+4 & 0.96-3.4 & \nodata & 25\\
J1813-1749 & 0.045 & 1.50E-13 & G12.8-0.02 & 4.6E+3 & 0.285-2.5 & \nodata & 26\\
B1823-13 & 0.101 & 7.53E-14 & GRS:J1825-137,PWN:G18.0-0.7 & 2.14E+4 & \nodata & 2 & 27\\
J1833-1034 & 0.062 & 2.02E-13 & G21.5-0.9 & 4.85E+3 & 0.72-1.07 & 1.857(6) & 28\\
J1841-0456 & 11.779 & 4.47E-11 & Kes 73 & 4.18E+3 & 1.1-1.5 & \nodata & 29\\
J1845-0256 & 6.971 & \nodata & G29.6+0.1 & 1.02E+8 & $\leqslant8$ & \nodata & 30\\
J1846-0258 & 0.326 & 7.08E-12 & Kes 75 & 728 & 0.9-4.3 & 2.65(1) & 31\\
J1850-0006 & 2.191 & 4.32E-15 & G32.45+0.1 & 8.04E+6 & \nodata & \nodata & \\
J1852+0040 & 0.105 & 8.68E-18 & Kes 79 & 1.92E+8 & 3-15 & \nodata & 32\\
B1853+01 & 0.267 & 2.08E-13 & W44 & 2.03E+4 & 6-29 & \nodata & 33\\
J1907+0602 & 0.107 & 8.68E-14 & G40.5-0.5? & 1.95E+4 & 20-40 & \nodata & 34\\
J1907+0919 & 5.169 & 7.78E-11 & G42.8+0.6 & 1.05E+3 & \nodata & \nodata & \\
J1930+1852 & 0.137 & 7.51E-13 & G54.1+0.3 & 2.89E+3 & 2.5-3.3 & \nodata & 35\\
B1951+32 & 0.040 & 5.84E-15 & CTB80 & 1.07E+5 & 60 & \nodata & 36\\
J1957+2831 & 0.308 & 3.11E-15 & G65.1+0.6 & 1.57E+6 & 44-140 & \nodata & 37\\
J2021+4026 & 0.265 & 5.47E-14 & G78.2+2.1 & 7.69E+4 & 6.6 & \nodata & 38\\
J2022+3842 & 0.024 & 4.32E-14 & G76.9+1.0 & 8.91E+3 & 5 & \nodata & 39\\
J2229+6114 & 0.052 & 7.83E-14 & G106.6+2.9 & 1.05E+4 & \nodata & \nodata & \\
J2301+5852 & 6.980 & 4.84E-13 & CTB109 & 2.28E+5 & 7.9-9.7 & \nodata & 40\\
B2334+61 & 0.495 & 1.93E-13 & G114.3+0.3 & 4.06E+4 & 7.7 & \nodata & 41\\
\enddata
%\tablecomments{The reference of SNR age and braking index is given bellow.}
\tablerefs{SNR ages:
[1] \citet{awd12,lht10}, [2] \citet{kot10,frh08}, [3] \citet{pbg03},
[4] \citet{aaa08}, [5] \citet{wg98}, [6] \citet{kp08,klh03}, [7] \citet{phs10,hph01}, [8] \citet{bpw12,gh09a},
[9] \citet{kp08,aet95} , [10] \citet{rm86,kp08}, [11] \citet{kp08,ksg12}, [12] \citet{kp08,wtr09}, [13] \citet{gh08},
[14] \citet{cks92}, [15] \citet{cng09}, [16] \citet{fz10a}, [17] \citet{njk96},
[19] \citet{fz10a}, [21] \citet{ckk87}, [22] \citet{rb02,vdg02}, [23] \citet{fo94}, [24] \citet{rb08},
[25] \citet{kp08,tr03,krv01}, [26] \citet{fz10b,gh09b}, [28] \citet{bb08,kp08}, [29] \citet{vk06},
[30] \citet{ggv99}, [31] \citet{kp08,gvb00}, [32] \citet{gsd09,sss04}, [33] \citet{kp08,kh95}, [34] \citet{dps80},
[35] \citet{kp08,bbg10}, [36] \citet{lr12,kp08}, [37] \citet{tl06}, [38] \citet{uta02}, [39] \citet{mcg11},
[40] \citet{spg04}, [41] \citet{yuk04}\\
Braking indices:
[4] \citet{lpg93}, [5] \citet{mmw06}, [7,16,31] \citet{lkg07}, [9,21,23,27] \citet{esp13},
[11] \citet{wje11}, [18] \citet{elk11}, [20] \citet{hgc09}, [28] \citet{rgl12}}
\end{deluxetable}

\end{document}